# One-dimensional Deep Low-rank and Sparse Network for Accelerated MRI

Zi Wang, Chen Qian, Di Guo, Hongwei Sun, Rushuai Li, Bo Zhao, and Xiaobo Qu*

***Abstract*—Deep learning has shown astonishing performance in accelerated magnetic resonance imaging (MRI). Most state-of-the-art deep learning reconstructions adopt the powerful convolutional neural network and perform 2D convolution since many magnetic resonance images or their corresponding k-space are in 2D. In this work, we present a new approach that explores the 1D convolution, making the deep network much easier to be trained and generalized. We further integrate the 1D convolution into the proposed deep network, named as One-dimensional Deep Low-rank and Sparse network (ODLS), which unrolls the iteration procedure of a low-rank and sparse reconstruction model. Extensive results on *in vivo* knee and brain datasets demonstrate that, the proposed ODLS is very suitable for the case of limited training subjects and provides improved reconstruction performance than state-of-the-art methods both visually and quantitatively. Additionally, ODLS also shows nice robustness to different undersampling scenarios and some mismatches between the training and test data. In summary, our work demonstrates that the 1D deep learning scheme is memory-efficient and robust in fast MRI.**

## I. INTRODUCTION

MAGNETIC resonance imaging (MRI) is a non-invasive imaging modality that enables visualization of anatomical structure and physiological functions[1]. However, its data acquisition process is relatively slow and imaging acceleration will improve patient comfort and reduce image artifacts[2]. Parallel imaging[3, 4] and sparse sampling[2, 5] have been employed to accelerate MRI through k-space undersampling.

Over the past two decades, many advanced methods have been established to reconstruct images from the undersampled data. They can be mainly categorized into two genres: Calibrated and non-calibrated methods. 1) Calibrated methods yield satisfactory results and commonly need autocalibration signals to estimate the coil sensitivity maps for encoding (e.g., SENSE[3]), or to estimate the weighting filters for k-space interpolation (e.g., GRAPPA[4] and SPIRiT[6]). Many insightful sparse[2, 7-14], low-rank[5, 15-17] priors or both of them[18-20] have been used to regularize the image or k-space for improved reconstructions. 2) Non-calibrated methods (e.g., structured low-rank[16, 17, 21-25]) may alleviate the dependence on autocalibration signals and lead to excellent reconstruction performances, arousing our interest to further develop them.

Recently, deep learning has shown great potential in accelerated MRI with the convolutional neural network (CNN)[26-28] and can be categorized into two classes[29]: End-to-end and unrolled networks. 1) For end-to-end networks, the direct inversion schemes without the data consistency have been employed[27, 30-32]. Although their performances are commendable, learning such large networks with many parameters requires large amount of training data, which may be scarce in some MRI applications. 2) The unrolled networks explore prior knowledges (e.g., sparsity and low-rankness), and unroll the iterative optimization procedures to deep neural networks. Most of them integrates sparse or other image-domain priors into the network design[26, 28, 33, 34], and they have been extended to parallel imaging through utilizing coil sensitivity maps[35-37]. In addition, the low-rank or other constraints have been extended in deep learning reconstructions, which show promising results[38-41].

As the image or k-space are usually in 2D, most state-of-the-art deep learning methods adopt 2D convolution even for 1D undersampling[32, 34, 35, 37, 40]. In this work, we present a new deep learning reconstruction method that employs 1D learning scheme, which is easier to train and more memory-efficient.

For the Cartesian 2D MRI, 1D phase encoding (PE) undersampling is commonly employed[2, 3, 42]. Along another dimension, the frequency encoding (FE), the k-space is fully-or non-sampled. Thus, as shown in Fig. 1, the undersampled 2D k-space can be split into many 1D hybrid spatial-phase encoding signals by taking the 1D FE inverse Fourier transform (IFT)[11]. Seizing this property, 1D reconstructions have shown improved performance than common 2D reconstructions in compressed sensing MRI[11].

This work was supported in part by the National Natural Science Foundation of China under grants 62122064, 61971361, 61871341, and 61811530021, the National Key R&D Program of China under grant 2017YFC0108703, the Natural Science Foundation of Fujian Province of China under grants 2021J011184, and the Xiamen University Nanqiang Outstanding Talents Program.

Zi Wang, Chen Qian, and Xiaobo Qu* are with the Department of Electronic Science, Biomedical Intelligent Cloud R&D Center, Fujian Provincial Key Laboratory of Plasma and Magnetic Resonance, National Institute for Data Science in Health and Medicine, Xiamen University, China (*Corresponding author, email: quxiaobo@xmu.edu.cn).

Di Guo is with the School of Computer and Information Engineering, Xiamen University of Technology, Xiamen, China.

Hongwei Sun is with the United Imaging Research Institute of Intelligent Imaging, Beijing, China.

Rushuai Li is with the Department of Nuclear Medicine, Nanjing First Hospital, Nanjing Medical University, Nanjing, China.

Bo Zhao is with the Department of Biomedical Engineering, Oden Institute for Computational Engineering and Sciences, University of Texas at Austin, USA.

For the deep learning, this 1D FE transform allows to reduce the dimension of reconstructed target signals from 2D to 1D, making the network training easier. Meanwhile, given the same number of available training subjects, the new scheme provides much more training samples, leading to more than two orders of the data augmentation compared to traditional 2D learnings. For example, a 2D image with the FE dimension of 256 can generate 256 1D training samples. Thus, 1D learning scheme is proposed to promote better network training and generalization.

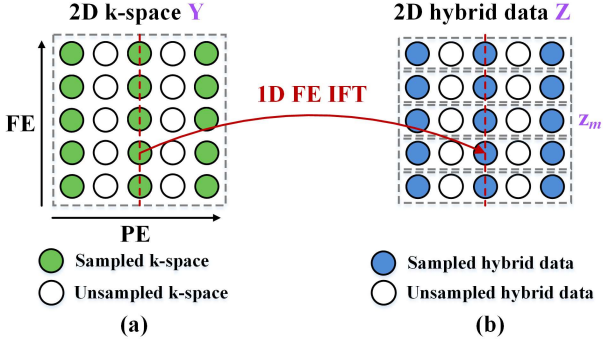

**Fig. 1.** Graphical illustration of splitting the 2D reconstruction problem into several 1D reconstructions. The region marked by a grey rectangle represents one of (a) the 2D reconstruction and (b) 1D reconstructions under the 1D PE undersampling. Note: The coil dimension is omitted.

Furthermore, inspired by the success of the reconstruction with sparse[11] and low-rank[17] priors, we integrate the 1D learning scheme with deep low-rank and sparse priors to develop a One-dimensional Deep Low-rank and Sparse network (ODLS). Specifically, we formulate both priors as regularized terms in a optimization model and unroll its iteration procedure into a deep network.

Our main contributions are summarized as follows:

1) A 1D learning scheme is proposed to make the deep network memory-efficient and easier to train, especially when the training subjects are limited.

2) A compact 1D CNN architecture is designed by unrolling the iteration procedure of a low-rank and sparse model, which explores the coil correlation and image sparsity, respectively.

3) Extensive results on *in vivo* datasets show that, ODLS is robust to different undersampling scenarios and some mismatches between the training and test data.

The remainder of this paper is organized as follows. Section II formulates the problem. Section III describes the proposed method and implementations. Section IV shows the results. Section V and VI provide the discussions and conclusion.

## II. PROBLEM FORMULATION

In the 1D PE undersampling, the 2D reconstruction can be split into several 1D reconstructions[11], as shown in Fig. 1. The specific processing procedure is as follows.

Consider a multi-coil 2D k-space $\mathbf{Y} \in \mathbb{C}^{M \times N \times J}$ with non-acquired positions zero-filled, which is acquired from the fully sampled multi-coil 2D k-space $\mathbf{X} \in \mathbb{C}^{M \times N \times J}$ by the undersampling operator $\mathcal{U}$. The acceleration factor (AF) of fast sampling is defined as $\mathrm{AF} = \dfrac{\text{fully sampled points}}{\text{undersampled points}}$.

Taking the 1D FE IFT $\mathcal{F}_{FE}^*$ of $\mathbf{Y}$, we can obtain the 2D hybrid data $\mathbf{Z} \in \mathbb{C}^{M \times N \times J}$:

$$\mathbf{Z} = \mathcal{F}_{FE}^* \mathbf{Y} = \mathcal{U}\mathcal{F}_{FE}^* \mathbf{X} = \mathcal{U}\mathbf{E}, \quad (1)$$

where $\mathbf{E} \in \mathbb{C}^{M \times N \times J}$ is the 2D hybrid data to be reconstructed. As the correlation between each row of $\mathbf{Y}$ has been decoupled by the 1D IFT, the $m^{\text{th}}$ row data $\mathbf{z}_m \in \mathbb{C}^{N \times J}$ of $\mathbf{Z}$ can be treated as an independent 1D hybrid signal. After reconstructing all rows (Fig. 1(b)), we can combine them sequentially along the FE to obtain $\hat{\mathbf{E}}$, and then take 1D PE IFT $\mathcal{F}_{PE}^*$ to yield the reconstructed image $\hat{\mathbf{S}} = \mathcal{F}_{PE}^* \hat{\mathbf{E}} \in \mathbb{C}^{M \times N \times J}$.

## III. PROPOSED METHOD

### A. One-dimensional Learning Scheme

Many existing deep learning methods adopt 2D learning schemes to train their 2D CNNs (Figs. 2(a)-(c)). For example, the 2D training schemes of the state-of-the-art methods IUNET[30], HDSLR[39], and DOTA[32] correspond to Figs. 2(a), 2(b), and 2(c), respectively. Different from them, our proposed 1D learning scheme (Fig. 2(d)) utilizes the 1D hybrid data as the training samples to train our 1D CNN.

Consider a dataset of $N_{TS}$ training subjects, each with $N_{slice}$ slices of size $M \times N \times J$. Table I shows that, by maximizing the use of available training subjects, our proposed 1D learning scheme owns the largest number of training samples and the smallest number of variables for per training sample. This nice property is conducive to the deep network training and generalization.

**TABLE I**
TRAINING INFORMATION FOR DIFFERENT LEARNING SCHEMES.

| Type of training samples | Conv. layer | Number of available training samples | Variables of each training sample |
|---|---|---|---|
| Fig. 2(a) | 2D | $N_{TS} \times N_{slice}$ | $M \times N \times J$ |
| Fig. 2(b) | | | |
| Fig. 2(c) | | | |
| Fig. 2(d) | 1D | $N_{TS} \times N_{slice} \times M$ | $N \times J$ |

### B. One-dimensional Low-rank and Sparse Model

In this work, we build a one-dimensional reconstruction model that introduces the low-rankness of multi-coil data and the sparsity of spatial signals in some transform domains.

The low-rank property of multi-coil data is achieved by constructing a cascaded Hankel matrix (Fig. 3). Here, we first structure multi-coil 1D hybrid data $\mathbf{e}_m \in \mathbb{C}^{N \times J}$, i.e., the $m^{\text{th}}$ row of $\mathbf{E}$, altogether into a cascaded Hankel as follows[16, 17, 21, 23, 43]:

$$\tilde{\mathcal{H}}(\mathbf{e}_m) = [\mathcal{H}(\mathbf{e}_{m,1}),...,\mathcal{H}(\mathbf{e}_{m,J})] \in \mathbb{C}^{G \times (N-G+1)J}, \quad (2)$$

where $\mathcal{H}$ is an operator that converts a vector into a Hankel matrix with a dimension of $G \times (N-G+1)$, and $G$ is the matrix pencil parameter. $\mathbf{e}_{m,J} \in \mathbb{C}^N$ is the $j^{\text{th}}$ coil $\mathbf{e}_m$. Assuming the coil sensitivities have compact k-space support, $\tilde{\mathcal{H}}(\mathbf{e}_m)$ enjoys low-rankness[21, 43, 44] since the effective rank

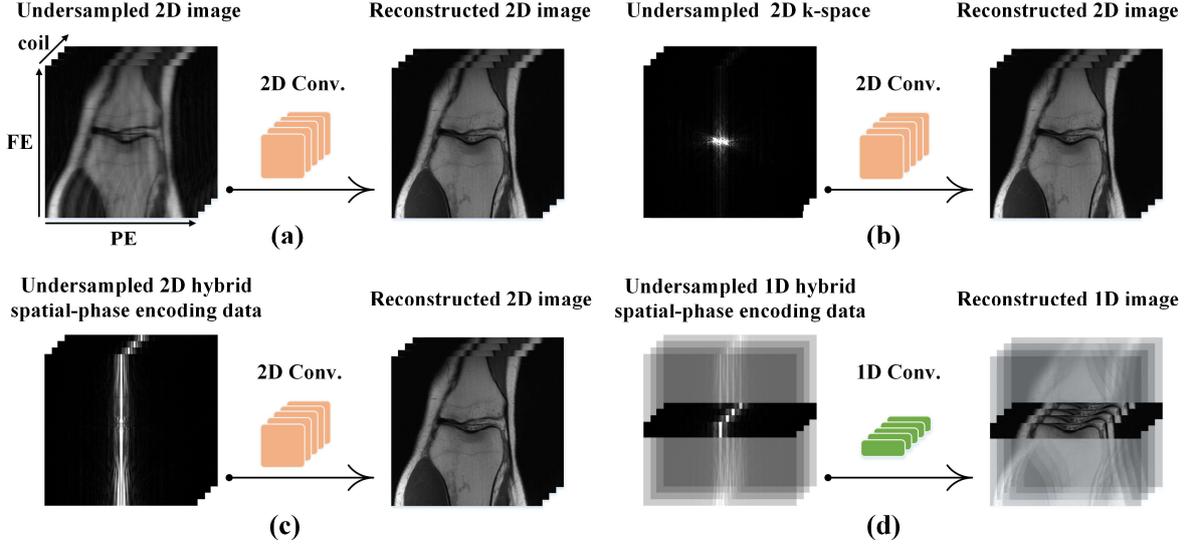

**Fig. 2.** Different CNN learning schemes for accelerated MRI. (a) 2D learning with image as training samples. (b) 2D learning with k-space as training samples. (c) 2D learning with 2D hybrid data as training samples. (d) The proposed 1D learning with 1D hybrid data as training samples. Note: "Conv." means the convolution layer.

of $\tilde{\mathcal{H}}(\mathbf{e}_m)$ can be much smaller than its size.

Then, each 1D signal can be reconstructed by enforcing the low-rankness of the cascaded Hankel matrix as

$$\min_{\mathbf{e}_m} \frac{1}{2}\|\mathbf{z}_m - \mathcal{U}\mathbf{e}_m\|_2^2 + \lambda_1 \left\|\tilde{\mathcal{H}}(\mathbf{e}_m)\right\|_*, \qquad (3)$$

where $\lambda_1$ is a regularization parameter.

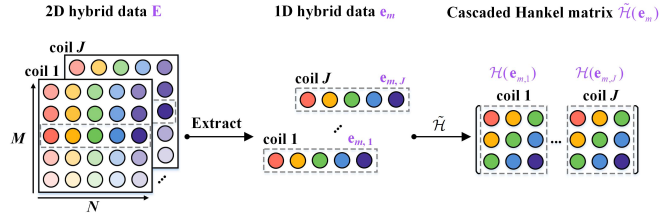

**Fig. 3.** Graphical illustration of constructing the Hankel matrix in our proposed method.

Another property that may improve the reconstruction is the sparsity of the 1D IFT of $\mathbf{e}_m$ in some transform domains[11]. By adding the $l_1$ norm into (3), the final 1D reconstruction can be formulated as

$$\min_{\mathbf{e}_m} \frac{1}{2}\|\mathbf{z}_m - \mathcal{U}\mathbf{e}_m\|_2^2 + \lambda_1 \left\|\tilde{\mathcal{H}}(\mathbf{e}_m)\right\|_* + \lambda_2 \left\|\mathcal{D}\mathcal{F}_{PE}^*\mathbf{e}_m\right\|_1, \qquad (4)$$

where $\mathcal{D}$ is the sparse transform, such as tight frames[12, 14], $\lambda_2$ is the regularization parameter for the $l_1$ norm.

To avoid time-consuming singular value decomposition, the iterative reweighted least squares scheme[24, 45] is adopted to majorize the nuclear norm with a weighted Frobenius norm as $\left\|\tilde{\mathcal{H}}(\mathbf{e}_m)\right\|_* \le \left\|\tilde{\mathcal{H}}(\mathbf{e}_m)\mathbf{Q}\right\|_F^2$, and (4) can be rewritten as

$$\min_{\mathbf{e}_m, \mathbf{Q}} \frac{1}{2}\|\mathbf{z}_m - \mathcal{U}\mathbf{e}_m\|_2^2 + \lambda_1 \left\|\tilde{\mathcal{H}}(\mathbf{e}_m)\mathbf{Q}\right\|_F^2 + \lambda_2 \left\|\mathcal{D}\mathcal{F}_{PE}\mathbf{e}_m\right\|_1, \qquad (5)$$

where $\mathbf{Q} \in \mathbb{C}^{BJ \times BJ}$ is a collection of column vectors spanning the null space of $\tilde{\mathcal{H}}(\mathbf{e}_m)$, and $B = N - G + 1$. (5) can be solved

by alternating sub-problems and the $k^{\text{th}}$ iteration is

$$\mathbf{e}_m^{(k)} = \arg\min_{\mathbf{e}_m} \frac{1}{2}\|\mathbf{z}_m - \mathcal{U}\mathbf{e}_m\|_2^2 + \lambda_1 \left\|\tilde{\mathcal{H}}(\mathbf{e}_m)\mathbf{Q}^{(k-1)}\right\|_F^2 + \lambda_2 \left\|\mathcal{D}\mathcal{F}_{PE}\mathbf{e}_m\right\|_1, \qquad (6)$$

$$\mathbf{Q}^{(k)} = \left[(\tilde{\mathcal{H}}(\mathbf{e}_m^{(k)}))^H (\tilde{\mathcal{H}}(\mathbf{e}_m^{(k)})) + \varepsilon\mathbf{I}\right]^{-\frac{1}{4}}, \qquad (7)$$

where $\mathbf{I}$ is the identity matrix. Consider the term $\tilde{\mathcal{H}}(\mathbf{e}_m)\mathbf{q}_b$, where $\mathbf{q}_b \in \mathbb{C}^{BJ}$ is the $b^{\text{th}}$ column of $\mathbf{Q}$. As $\mathcal{H}(\mathbf{e}_{m,j})$ is a Hankel matrix, $\mathcal{H}(\mathbf{e}_{m,j})\mathbf{q}_{b,1}$ represents the linear convolution between $\mathbf{e}_{m,j}$ and $\mathbf{q}_{b,1}$. Since the convolution is commutative,

$$\tilde{\mathcal{H}}(\mathbf{e}_m)\mathbf{q}_b = [\mathcal{H}(\mathbf{e}_{m,1}),...,\mathcal{H}(\mathbf{e}_{m,J})]\begin{bmatrix}\mathbf{q}_{b,1}\\ \vdots \\ \mathbf{q}_{b,J}\end{bmatrix}$$

$$= [\mathcal{T}(\mathbf{q}_{b,1}),...,\mathcal{T}(\mathbf{q}_{b,J})]\begin{bmatrix}\mathbf{e}_{m,1}\\ \vdots \\ \mathbf{e}_{m,J}\end{bmatrix} = \tilde{\mathcal{T}}(\mathbf{q}_b)\mathbf{e}_m, \qquad (8)$$

where $\tilde{\mathcal{T}}(\mathbf{q}_b)$ is a cascaded Toeplitz matrix constructed from the elements of $\mathbf{q}_b$. Thus, we have[39]

$$\left\|\tilde{\mathcal{H}}(\mathbf{e}_m)\mathbf{Q}\right\|_F^2 = \left\|\tilde{\mathcal{P}}(\mathbf{Q})\mathbf{e}_m\right\|_F^2, \qquad (9)$$

where $\tilde{\mathcal{P}}(\mathbf{Q})$ is arranged by vertically cascading $\tilde{\mathcal{T}}(\mathbf{q}_b)$.

Using the relationship in (9), then (6) can be rewritten as

$$\mathbf{e}_m^{(k)} = \arg\min_{\mathbf{e}_m} \frac{1}{2}\|\mathbf{z}_m - \mathcal{U}\mathbf{e}_m\|_2^2 + \lambda_1 \left\|\tilde{\mathcal{P}}(\mathbf{Q}^{(k-1)})\mathbf{e}_m\right\|_F^2 + \lambda_2 \left\|\mathcal{D}\mathcal{F}_{PE}\mathbf{e}_m\right\|_1. \qquad (10)$$

To solve (10), projected iterative soft-thresholding algorithm [12, 37] is chosen to enable the iterative procedure:

$$\begin{cases}\mathbf{r}_m^{(k)} = \lambda_1[(\tilde{\mathcal{P}}(\mathbf{Q}^{(k-1)}))^H (\tilde{\mathcal{P}}(\mathbf{Q}^{(k-1)})\mathbf{e}_m^{(k-1)})]\\ \mathbf{d}_m^{(k)} = \mathbf{e}_m^{(k-1)} - \gamma[\mathcal{U}^*(\mathcal{U}\mathbf{e}_m^{(k-1)} - \mathbf{z}_m) + 2\mathbf{r}_m^{(k)}],\\ \mathbf{e}_m^{(k)} = \mathcal{F}_{PE}\mathcal{D}^*[soft(\mathcal{D}\mathcal{F}_{PE}^*\mathbf{d}_m^{(k)};\gamma\lambda_2)]\end{cases} \qquad (11)$$

and $\mathbf{Q}$ is updated as shown in (7) at each iteration. $\gamma$ is the step size, $soft(x; \rho) = \max\{|x| - \rho\} \cdot x/|x|$ is the element-wise soft-thresholding, and the superscript $*$ represents the adjoint operation. Specifically, the sparse transform satisfies $\mathcal{D}^*\mathcal{D} = \mathcal{I}$, where $\mathcal{I}$ is the identity transform.

Conventional reconstruction iteratively solves each sub-problem in (11) until the algorithm converges. Thus the whole procedure may cost long reconstruction time[19]. Moreover, all parameters $\{\lambda_1, \lambda_2, \gamma\}$ and the sparse transform $\mathcal{D}$ has to be manually selected, which is cumbersome and may not be robust for different scenarios.

### C. ODLS: One-dimensional Deep Low-rank and Sparse Network

Here, we propose a One-dimensional Deep Low-rank and Sparse network (ODLS) by unrolling the iteration procedure in (11) to a deep learning scheme. Specifically, three sub-problems in (11) correspond to three network modules (Fig. 4). In ODLS, parameters $\{\lambda_1, \lambda_2, \gamma\}$, the sparse transform $\mathcal{D}$, and the filterbanks $\mathbf{Q}$ are learnable.

The detailed description of three modules are as follows.

#### 1) Deep low-rank module

Fig. 4(b) is the deep low-rank module that corresponds to the first sub-equation of (11).

$(\tilde{\mathcal{P}}(\mathbf{Q}))^H (\tilde{\mathcal{P}}(\mathbf{Q})\mathbf{e}_m)$ can be regarded as feeding $\mathbf{e}_m$ into multi-input-multi-output filterbanks which are determined by $\mathbf{Q}$ [39]. Here, we pre-train CNN filterbanks to replace $\mathbf{Q}$ due to the strong learning ability of the network. It adopts the similar idea of calibrated methods[4, 6, 16, 43]. The common goal is to estimate the proper null space and significantly reduce the computational complexity. The difference is that they estimate $\mathbf{Q}$ from autocalibration signals, whereas we obtain filterbanks from enormous training datasets.

In this paper, we use the multi-layer encoder-decoder 1D CNN in $k^{th}$ network phase to replace $\mathbf{Q}^{(k-1)}$, The deep low-rank module is then designed as

$$\mathbf{r}_m^{(k)} = \lambda_1^{(k)}[\mathcal{N}_1(\mathbf{e}_m^{(k-1)})], \qquad (12)$$

where $\mathcal{N}_1$ is a multi-layer encoder-decoder 1D CNN. $\lambda_1^{(k)}$ sets as a learnable parameter initialized to 0.001 and is changed at each network phase. When $k = 1$, the initialized network input $\mathbf{e}_m^{(0)} = \mathcal{U}^*\mathbf{z}_m$ is the zero-filled hybrid data that has strong artifacts.

#### 2) Data consistency module

In this module, each output is forced to ensure the reconstructed k-space are aligned to acquired data (Fig. 4(c)). The data consistency module is designed mostly same to the second sub-problem of (11). The only difference is that we set $\gamma$ as a learnable parameter initialized to 1. The data consistency module is denoted as

$$\mathbf{d}_m^{(k)} = \mathbf{e}_m^{(k)} - \gamma^{(k)}[\mathcal{U}^*(\mathcal{U}\mathbf{e}_m^{(k-1)} - \mathbf{z}_m) + 2\mathbf{r}_m^{(k)}], \qquad (13)$$

#### 3) Deep sparse module

Fig. 4(c) is the deep sparse module that corresponds to the third sub-function of (11).

Here, we replace forward transform $\mathcal{D}$ and inverse transform $\mathcal{D}^*$ with two multi-layer 1D CNNs $\mathcal{N}_2$ and $\mathcal{N}_3$, respectively, to learn the general sparse transform from training datasets. Ideally, we hope the sparse transform being invertible, i.e., $\mathcal{D}^*\mathcal{D} = \mathcal{I}$. However, it may be not satisfied well in network learning. Thus, we further integrating the term $\|\mathcal{N}_3(\mathcal{N}_2(x)) - x\|_2^2$ into the total loss function (15). Finally, the deep sparse module is designed as

$$\mathbf{e}_m^{(k)} = \mathcal{F}_{PE}\mathcal{N}_3[soft(\mathcal{N}_2(\mathcal{F}_{PE}^*\mathbf{d}_m^{(k)}); \theta^{(k)})], \qquad (14)$$

where $\theta^{(k)}$ is the learnable threshold initialized to 0.001 and we allow it to vary at each network phase.

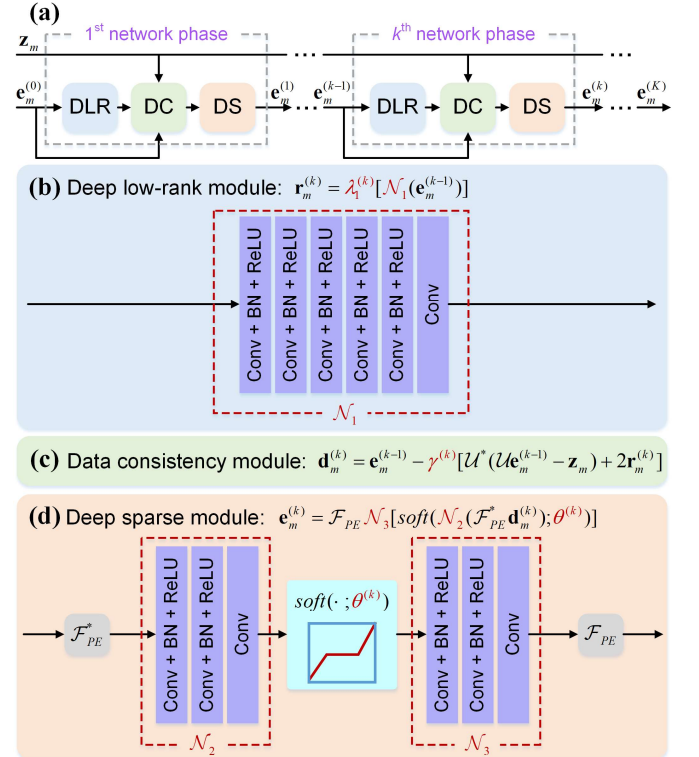

Fig. 4. The proposed ODLS for accelerated MRI. (a) The recursive ODLS architecture that alternates between the deep low-rank (DLR) module, data consistency (DC) module, and deep sparse (DS) module. (b)-(d) The detailed structures of the DLR, DC, and DS modules, respectively. Note: The variables marked red are learnable. "BN" is the batch normalization

#### 4) Ablation study

To show the effectively of using both low-rank and sparse priors, we compare the proposed method with the two methods that only impose 1D low-rank or sparse priors, named as ODLS-L and ODLS-S respectively.

As shown in Fig. 5, by using only one prior, both ODLS-L and ODLS-S recover most of the image structure but still remains some artifacts. The ODLS, which equipped with both low-rank and sparse priors, provides the image with lowest error (RLNE=0.0734) and highest similarity (PSNR=36.53dB, SSIM=0.9352). These results demonstrate the effectiveness of using two complementary information in MRI reconstruction.

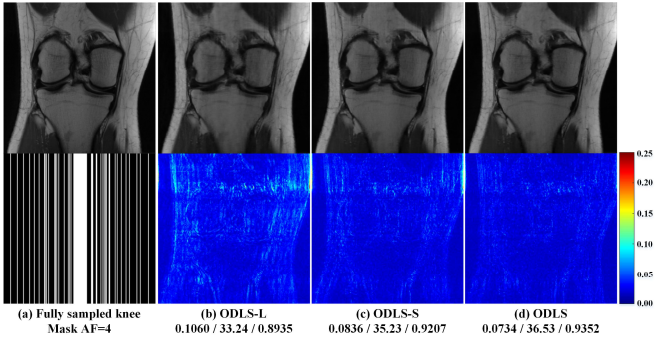

**Fig. 5.** Ablation study of knee reconstructions. (a) is the fully sampled image and the Cartesian undersampling pattern with AF=4. (b)-(d) are reconstructed images and the corresponding error maps. Note: RLNE/PSNR(dB)/SSIM are listed for each reconstruction.

It is worth noting that the proposed ODLS is totally different from the SLR-Net[41], which is a state-of-the-art deep low-rank and sparse network. The main differences are as follows: 1) SLR-Net requires traditional singular value decomposition and only the threshold of singular values is learnable. The proposed ODLS avoids the singular value decomposition since it is time-consuming for a large-scale constrained matrices. 2) SLR-Net is designed for dynamic MRI, whereas the proposed ODLS focuses on static MRI. 3) SLR-Net constrains the 3D spatiotemporal data to build a 3D CNN, while we constructs a 1D CNN through constraining the 1D hybrid data.

### D. Network Architecture and Loss Function

As shown in Fig. 4, the proposed 1D CNN-based ODLS is an unrolled recursive network. $K = 10$ is an optimal trade-off between the reconstruction performance and time consumption. At each network phase, $\mathcal{N}_1$, $\mathcal{N}_2$, and $\mathcal{N}_3$ consist of 6, 3, and 3 convolutional layers, respectively. Each convolutional layer contains 48 1D convolution filers of size 3, followed by batch normalization[46] and an ReLU as activation function.

To demonstrate the effectiveness of 1D learning of the proposed ODLS, we also propose a 2D CNN-based TDLS (two-dimensional deep low-rank and sparse network) for comparison. It has the similar network architecture to ODLS, but each convolutional layer contains 48 2D convolution filers of size 3×3, followed by BN and ReLU.

In the training stage, ODLS and TDLS obeyed the schemes in Fig. 2(d) and Fig. 2(b), respectively. The network weights of both ODLS and TDLS were Xavier initialized[47] and trained for 300 epochs with the Adam optimizer. Their initial learning rate was set to 0.001 with an exponential decay of 0.99. The batch size of ODLS and TDLS was 128 and 2, respectively. The loss functions of ODLS and TDLS are defined as follows:

$$\mathcal{L}_{total}^{ODLS}(\mathbf{\Theta}^{ODLS}) = \mathcal{L}_{err}^{ODLS} + \beta \mathcal{L}_{sym}^{ODLS}$$
$$= \frac{1}{KT}\sum_{k=1}^{K}\sum_{t=1}^{T}\left(\left\|\mathcal{F}_{PE}(\mathbf{e}_m^{ref,t} - \mathbf{e}_m^{(k),t})\right\|_2^2 + \beta\left\|\mathcal{N}_3(\mathcal{N}_2(\mathcal{F}_{PE}\mathbf{e}_m^{(k),t})) - \mathcal{F}_{PE}^*\mathbf{e}_m^{(k),t}\right\|_2^2\right),$$
$$(15)$$

$$\mathcal{L}_{total}^{TDLS}(\mathbf{\Theta}^{TDLS}) = \mathcal{L}_{err}^{TDLS} + \beta \mathcal{L}_{sym}^{TDLS}$$
$$= \frac{1}{KT}\sum_{k=1}^{K}\sum_{t=1}^{T}\left(\left\|\mathcal{F}_{2D}(\mathbf{X}^{ref,t} - \mathbf{X}^{(k),t})\right\|_2^2 + \beta\left\|\mathcal{N}_3(\mathcal{N}_2(\mathcal{F}_{2D}^*\mathbf{X}^{(k),t})) - \mathcal{F}_{2D}^*\mathbf{X}^{(k),t}\right\|_2^2\right)$$
$$(16)$$

where $T$ is the number of training samples (different in ODLS and TDLS), $\beta$ is set as 0.01. $\mathcal{L}_{sym}^{ODLS}$ and $\mathcal{L}_{sym}^{TDLS}$ are symmetry loss terms of inversion in ODLS and TDLS, respectively. $\mathbf{e}_m^{ref,t}$ and $\mathbf{X}^{ref,t}$ are the labels of the $t^{th}$ training sample of ODLS and TDLS, respectively. The proposed networks were implemented on a server equipped with dual Intel Xeon Silver 4210 CPUs, 128 GB RAM, and the Nvidia Tesla T4 GPU (16 GB memory) in Tensorflow 1.15.0[48]. The training of ODLS and TDLS took about 5 hours and 8 hours, respectively.

In the reconstruction stage, for given undersampled k-space data, we can reconstruct them through the trained ODLS and TDLS, respectively. The reconstruction process of TDLS is similar to many existing 2D CNN-based methods, while that of the 1D CNN-based ODLS is special. As shown in Fig. 6, for ODLS, the 1D FE IFT is first performed on the undersampled k-space to obtain $\mathbf{Z} = \mathcal{F}_{FE}^*\mathbf{Y}$, all rows of $\mathbf{Z}$ form a batch that is then reconstructed in parallel and stitched back together to yield the reconstructed hybrid data $\hat{\mathbf{E}}$. After performing the 1D PE IFT, we can obtain the final reconstructed image $\hat{\mathbf{S}} = \mathcal{F}_{PE}^*\hat{\mathbf{E}}$.

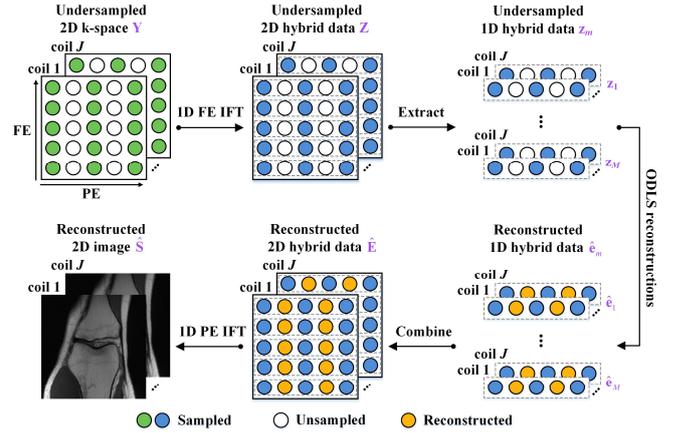

**Fig. 6.** Flowchart of the 1D reconstructions using the proposed ODLS.

## IV. EXPERIMENTAL RESULTS

### A. Datasets

Four datasets were used in this paper: Two knee datasets from an open repository[35] and two brain datasets from our own collection. Their study protocol was Institutional Review Board approved and informed consent was obtained from volunteers before examination. The detailed acquisition specifications for the datasets are provided below.

#### 1) Knee datasets

Two knee datasets are coronal and sagittal proton density weighted k-space data. They were acquired from 2D turbo-spin-echo sequence at a 3T MRI system with 15 coils (Siemens Magnetom Skyra).

For the coronal proton density weighted dataset: TR=2750 ms, TE=27 ms, matrix size 320×288. There are 20 subjects and each subject contains about 35 slices of size 640×368×15. Twelve subjects were used for training, one for validation, and the remaining for test.

For the sagittal proton density weighted dataset: TR=2800

ms, TE=27 ms, matrix size 384×307. There are 8 subjects and each subject contains about 35 slices of size 768×770×15. All subjects were used for the mismatched reconstruction.

### 2) Brain datasets

Two brain datasets are axial $T_2$ weighted and $T_1$ weighted k-space data. The experiments were conducted on a 3T MRI system with 32 coils (United Imaging uPMR 790) using 2D fast-spin-echo sequence.

For the axial $T_2$ weighted dataset: TR=8091 ms, TE=112.8 ms, matrix size 224×225, average time 2. There are 8 subjects and each subject contains about 70 slices of size 448×225×32. Six subjects were used for training, one for validation, and the remaining for test.

For the axial $T_1$ weighted dataset: TR=2234 ms, TE=12.5 ms, matrix size 224×225, average time 2. There are 2 subjects and each subject contains about 70 slices of size 448×225×32. All subjects were used for the mismatched reconstruction.

To reduce computational complexity, coil compression[49] was used to decrease the number of coils to 8 in all dataset, and the image size were center-cropped to 224×224. All datasets were fully sampled, and they were first retrospectively undersampled, then used for training and test. To increase the generalization of the network, in the same undersampling scenario, each image of training and test datasets owns the different undersampling masks with the same AF.

### B. Compared Methods and Evaluation Criteria

For comparative study, conventional parallel imaging method GRAPPA[4] was used as the reconstruction baseline. We also compared the proposed 1D CNN-based ODLS with the proposed 2D CNN-based TDLS, and three state-of-the-art 2D CNN-based methods including IUNET[30], HDSLR[39], and DOTA[32]. The training schemes of IUNET, HDSLR, and DOTA are representative, as shown in Figs. 2(a)-(c), respectively. Both IUNET and DOTA own the non-unrolled network architectures, and DOTA further introduces the fully-connected layers to replace the 1D IFT, whereas HDSLR is an unrolled network using joint low-rank and image-domain priors. All deep learning methods were executed according to the typical setting mentioned by the authors. Parameters of GRAPPA were optimized to obtain the lowest RLNE (See Supplement S4).

Notably, by virtue of 1D CNN, the proposed ODLS has few trainable parameters (664350), about 3% of IUNET, 2% of DOTA, 26% of HDSLR, and 34% of TDLS, thus it is memory-efficient.

All methods were implemented on a server equipped with dual Intel Xeon Silver 4210 CPUs, 128 GB RAM, and the Nvidia Tesla T4 GPU (16 GB memory). The typical reconstruction time per slice are GRAPPA (8.71 seconds), IUNET (0.07 seconds), DOTA (0.11 seconds), HDSLR (0.14 seconds), TDLS (0.13 seconds), and ODLS (0.10 seconds).

In all experiments, the reconstructed multi-coil images were finally displayed after combining by the square root of sum of squares.

To quantitatively evaluate the reconstruction performance, we utilized three evaluation criteria (See definitions in Supplement S1): The relative $l_2$ norm error (RLNE)[10] as the error-measure, the peak signal-to-noise ratio (PSNR) and structural similarity index (SSIM)[50] as the similarity measures. The lower RLNE, higher PSNR, and higher SSIM indicate the lower reconstruction error, less image distortions, and better detail preservation in reconstructions, respectively.

### C. 1D versus 2D Learning with Limited Training Subjects

Here, we trained networks on a knee dataset of different number of training subjects $N_{TS}$, to compare the reconstruction performance of the 2D learning and the proposed 1D learning scheme. First, to eliminate interferences of the network design, the propose network TDLS and ODLS which only differed in the size of convolution filers (2D versus 1D) were used.

As shown in Fig. 7(a), for 2D learning TDLS, we can observe that the gap between the training and validation loss decreases with the increase of $N_{TS}$. For $N_{TS}$=12, the gap is small enough to imply that the network has been properly trained. But when the network is trained using low data availability ($N_{TS}$=2,4), the 2D learning is poor. For 1D learning ODLS in Fig. 7(b), the gaps between the training and validation loss are smaller, and the convergence speeds are faster, than those of the 2D learning. Moreover, the convergence speed of ODLS is highly independent on $N_{TS}$. And starting from $N_{TS}$=4, its gaps become small which indicates good training and generalization.

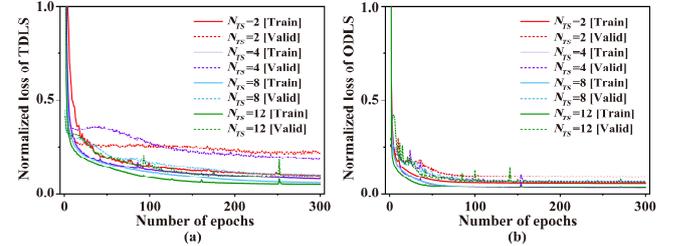

Fig. 7. The loss curves of 2D and 1D learning for different number of training subjects $N_{TS}$. (a) 2D learning TDLS. (b) 1D learning ODLS. Note: Solid and dashed lines indicate the loss function of the training and validation, respectively. [Train]: training stage, [Valid]: validation stage.

TABLE II
RLNE (×10²)/PSNR (dB)/SSIM (×10²) OF KNEE RECONSTRUCTIONS USING DIFFERENT NUMBER OF TRAINING SUBJECTS (MEAN±STD).

| Method | $N_{TS}$ | RLNE | PSNR | SSIM |
|---|---|---|---|---|
| GRAPPA | / | 15.46±2.13 | 30.44±1.31 | 83.27±2.59 |
| 2D learning TDLS | 2 | 19.96±2.66 | 27.56±1.49 | 79.26±3.68 |
| | 4 | 16.04±2.28 | 29.76±1.55 | 83.99±3.12 |
| | 8 | 11.77±1.92 | 32.53±1.71 | 88.62±2.23 |
| | 12 | 10.76±1.60 | 33.39±1.66 | 89.75±1.99 |
| 1D learning ODLS | 2 | 14.71±2.68 | 30.01±1.96 | 83.93±3.57 |
| | 4 | 11.87±1.91 | 32.07±1.79 | 87.36±2.47 |
| | 8 | 10.46±1.71 | 33.37±1.77 | 89.06±2.14 |
| | 12 | 9.78±1.54 | 34.15±1.81 | 90.02±2.01 |

Note: The Cartesian undersampling pattern with AF=4 is used. "/" represents that the corresponding method requires no training. The means and standard deviations are computed over all test data, respectively.

The quantitative comparison in Table II shows that, for any $N_{TS}$, the proposed 1D learning consistently outperforms the 2D learning, even the latter has been properly trained ($N_{TS}$=8,12). We also observe that, even for $N_{TS}$=2, the 1D learning can already provide slightly better reconstructions than the baseline

GRAPPA in the terms of RLNE and SSIM, while the 2D learning starts to show its advantage only when $N_{TS}$=8. Moreover, the results of the 1D learning ($N_{TS}$=8) are already comparable to the 2D learning ($N_{TS}$=12).

These results imply that, the proposed 1D learning scheme is suitable for training a network with limited subjects.

### D. Comparison with State-of-the-art Methods

To further validate the advantages of 1D learning with limited training subjects, the comparison with other state-of-the-art 2D learning networks was carried out.

Figs. 8(a)-(c) show that, for any number of training subjects $N_{TS}$, the proposed 1D learning ODLS consistently outperforms other 2D learning networks in terms of RLNE, PSNR, and SSIM. We can observe that ODLS can already provide better reconstructions than the baseline GRAPPA in terms of RLNE and SSIM when training subjects are very limited ($N_{TS}$=2). Differently, 2D learning IUNET, DOTA, and HDSLR start to outperform the baseline GRAPPA when $N_{TS}$≥4. This is because the 2D learning methods always needs a large number of training subjects to make the network sufficiently trained. The similar phenomena can also be found in the reconstruction of a brain dataset (Figs. 8(d)-(f)). These results imply that, the proposed method is more useful when available training subjects are very limited.

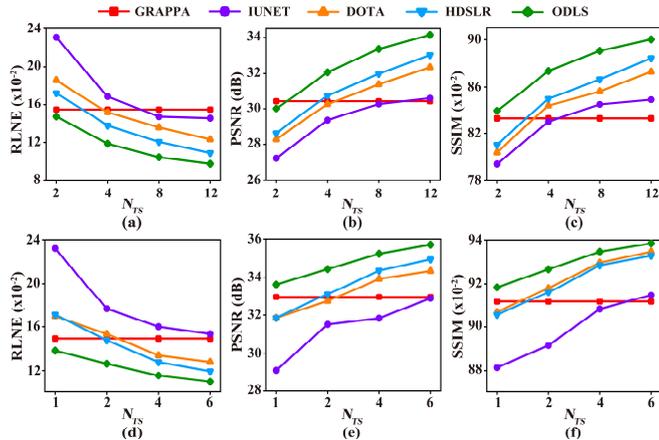

**Fig. 8.** Quantitative comparison of reconstructions using different number of training subjects $N_{TS}$ of a knee and a brain dataset. (a)-(c) are the mean of RLNE, PSNR, SSIM of the reconstructions of the coronal proton density weighted knee dataset, respectively. (d)-(f) are the mean of RLNE, PSNR, SSIM of the reconstructions of the axial $T_2$ weighted brain dataset, respectively. Note: The Cartesian undersampling pattern with AF=4 is used. The means are computed over all test data, and the standard deviations can be found in Supplement Table S2-1 and S2-2.

Representative images of two datasets are shown in Fig. 9. In the reconstruction under the Cartesian undersampling pattern, GRAPPA (Figs. 9(b)(h)), IUNET (Figs. 9(c)(i)), and DOTA (Figs. 9(d)(j)) yield results exhibiting obvious artifacts. Both HDSLR (Figs. 9(e)(k)) and the proposed ODLS (Figs. 9(f)(l)) provide the images with nice artifacts suppression. However, the reconstruction errors of HDSLR appear larger than that of ODLS, especially in the knee edge and brain skull area. Meanwhile, ODLS outperforms other state-of-the-art methods in terms of three evaluation criteria, indicating the good ability of the artifacts suppression and detail preservation.

## V. DISCUSSIONS

### A. Mismatched Reconstruction

Mismatched reconstruction refers to utilizing a trained network to reconstruct the test dataset which is different from the acquisition specification of the training dataset. In this part, we focused on the mismatch of the knee plane orientation and brain contrast weighting, respectively.

For the mismatch of the knee plane orientation, the networks trained by the coronal dataset were used to reconstruct the unseen sagittal one. Figs. 10(a)-(f) show that, the proposed ODLS yields the lowest reconstruction errors, while other four state-of-the-art methods provide the images with obvious artifacts. The statistical quantitative comparisons are listed in Table III, showing that ODLS not only outperforms other deep learning methods, but also is the only one that surpasses the baseline GRAPPA on all evaluation criteria.

TABLE III
RLNE (×$10^{-2}$)/PSNR (dB)/SSIM (×$10^{-2}$) OF RECONSTRUCTIONS UNDER MISMATCH (MEAN±STD).

| Mismatch | Method | RLNE | PSNR | SSIM |
|---|---|---|---|---|
| Knee plane orientation: coronal → sagittal | GRAPPA | 13.55±2.18 | 31.53±1.71 | 85.05±3.06 |
| | IUNET | 15.35±1.32 | 29.32±1.31 | 82.13±2.69 |
| | DOTA | 13.61±1.56 | 30.78±1.45 | 85.01±2.71 |
| | HDSLR | 13.15±1.74 | 30.86±1.41 | 84.53±2.79 |
| | ODLS | **11.07±1.37** | **32.33±1.63** | **87.42±2.33** |
| Brain contrast weighting: $T_2$ → $T_1$ | GRAPPA | 13.92±2.63 | 28.72±2.24 | 91.05±3.08 |
| | IUNET | 17.20±2.69 | 26.83±1.99 | 87.52±2.90 |
| | DOTA | 13.18±2.45 | 29.18±1.93 | 92.46±2.06 |
| | HDSLR | 12.40±2.92 | 29.79±2.17 | 92.37±2.28 |
| | ODLS | **10.94±2.24** | **30.83±2.16** | **93.31±2.17** |

Note: The Cartesian undersampling pattern with AF=4 is used. In A→B, A is the training dataset and B is the test dataset. The deep learning methods are trained using all available training subjects. The lowest RLNE, highest PSNR and SSIM values are bold faced.

The similar robust results of ODLS can also be found for the mismatch of the brain contrast weighting. In this experiment, we trained networks by the $T_2$ weighted dataset to reconstruct the unseen $T_1$ weighted one. The results in Figs. 10(g)-(l) and Table III show that, ODLS provides better reconstructions than other four methods visually and quantitatively.

These results demonstrate that, comparing with other deep learning methods, the proposed ODLS has improved generalization capabilities, and can also provide better artifacts suppression than GRAPPA.

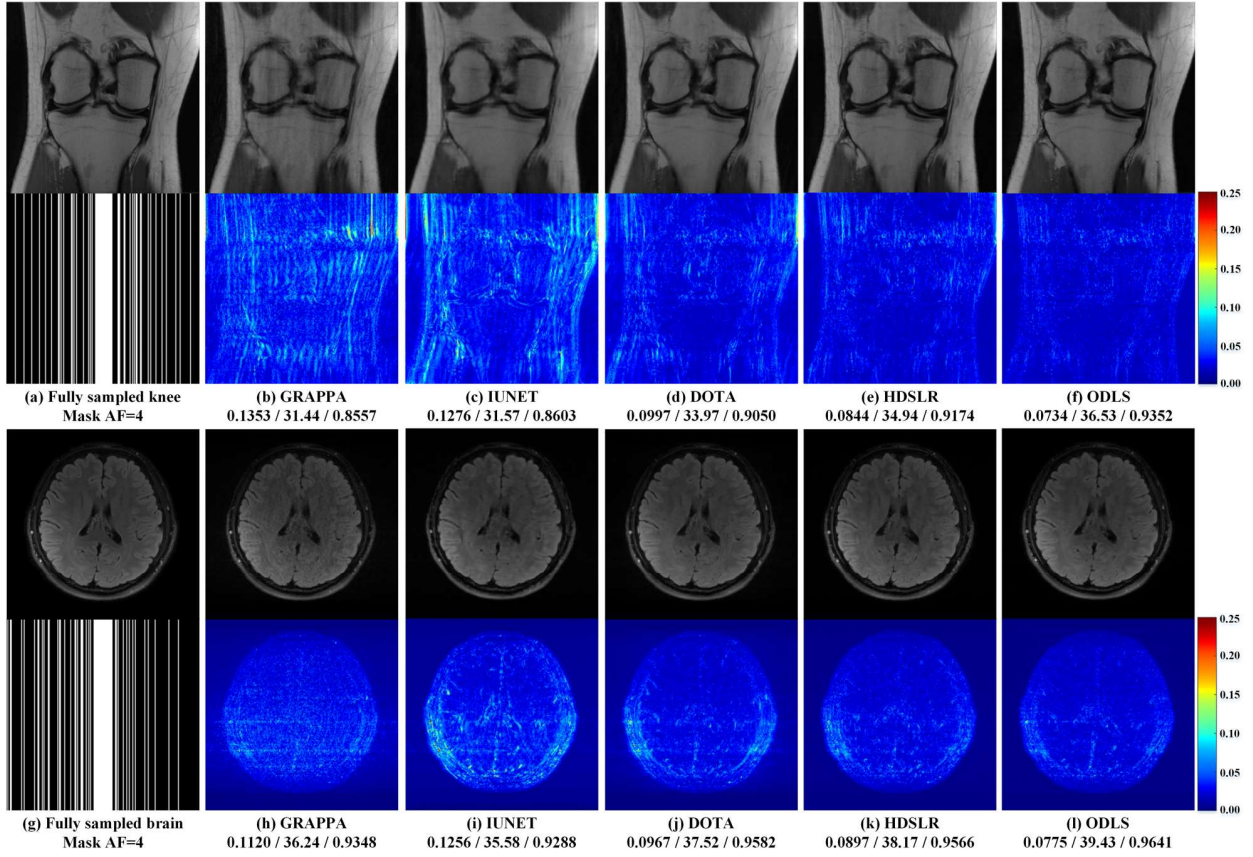

Fig. 9. Reconstruction results of knee and brain datasets using different methods. (a) (or (g)) is the fully sampled image and the Cartesian undersampling pattern with AF=4. (b)-(f) (or (h)-(l)) are reconstructed images and the corresponding error maps. Note: RLNE/PSNR(dB)/SSIM are listed for each reconstruction. The deep learning methods are trained using all available training subjects.

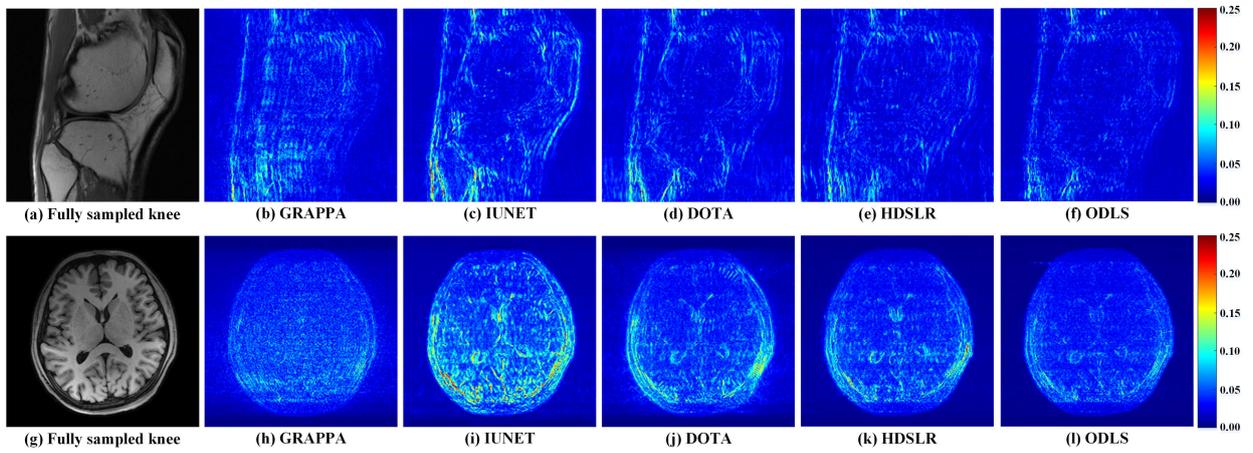

Fig. 10. Mismatched reconstruction results of different methods under the Cartesian undersampling patterns with AF=4. (a) (or (g)) is the fully sampled image. (b)-(f) (or (h)-(l)) are reconstruction error maps.

## B. Reconstruction under Different Undersampling Scenarios

To further evaluate the proposed ODLS, we reconstructed a knee dataset and a brain dataset under different undersampling scenarios, including the Cartesian undersampling pattern with AF=6 (Fig. 11(a)), the 3/4 partial Fourier undersampling pattern with AF=3 (Fig. 11(b)), and the uniform undersampling patterns with AF=4 (Fig. 11(c)). The state-of-the-art methods, including GRAPPA and an unrolled model-based deep learning method HDSLR, were taken for comparison.

The statistical quantitative comparison and representative reconstruction results under different undersampling scenarios can be found in Table IV, Supplement Fig. S3-1 and S3-2. Compared with GRAPPA and HDSLR, the proposed ODLS achieves better artifacts suppression and detail preservation regardless of the undersampling scenario, confirmed by lower RLNE, higher PSNR, and higher SSIM. Thus, we conclude that ODLS is applicable to different undersampling scenarios.

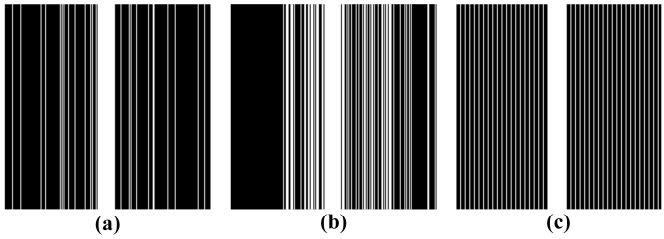

**Fig. 11.** Three representative undersampling patterns. (a) The Cartesian undersampling pattern with AF=6. (b) The 3/4 partial Fourier undersampling pattern with AF=3. (c) The uniform undersampling pattern with AF=4.

TABLE IV

RLNE (×10⁻²)/PSNR (dB)/SSIM (×10⁻²) OF RECONSTRUCTIONS UNDER DIFFERENT UNDERSAMPLING SCENARIOS (MEAN±STD).

| Coronal proton density weighted knee dataset | | | |
|---|---|---|---|
| Pattern | Method | RLNE | PSNR | SSIM |
| Cartesian AF=6 | GRAPPA | 20.29±2.80 | 27.35±1.53 | 76.32±3.77 |
| | HDSLR | 14.35±2.45 | 29.81±1.89 | 82.43±3.26 |
| | ODLS | **13.02±2.08** | **30.57±1.86** | **84.55±3.25** |
| 3/4 partial Fourier AF=3 | GRAPPA | 11.82±1.80 | 33.37±1.67 | 89.14±2.28 |
| | HDSLR | 9.17±1.22 | 35.37±1.77 | 92.13±1.66 |
| | ODLS | **8.36±0.92** | **36.32±1.70** | **93.21±1.61** |
| Uniform AF=4 | GRAPPA | 16.68±1.87 | 29.90±1.36 | 81.90±2.58 |
| | HDSLR | 11.57±1.81 | 32.33±1.78 | 87.67±2.18 |
| | ODLS | **10.11±1.35** | **33.60±1.61** | **89.55±1.68** |
| Axial T₂ weighted brain dataset | | | |
| Pattern | Method | RLNE | PSNR | SSIM |
| Cartesian AF=6 | GRAPPA | 18.68±1.79 | 30.98±1.64 | 87.76±3.46 |
| | HDSLR | 16.70±2.11 | 31.98±1.58 | 88.68±3.21 |
| | ODLS | **15.28±1.91** | **32.76±1.60** | **89.85±3.06** |
| 3/4 partial Fourier AF=3 | GRAPPA | 12.40±1.71 | 34.57±1.78 | 93.59±2.22 |
| | HDSLR | 10.73±1.95 | 35.88±1.82 | 94.92±1.88 |
| | ODLS | **9.97±1.76** | **36.52±1.87** | **95.15±1.91** |
| Uniform AF=4 | GRAPPA | 15.99±1.84 | 32.35±1.60 | 90.99±2.84 |
| | HDSLR | 11.20±1.95 | 35.51±1.72 | 93.93±1.96 |
| | ODLS | **10.44±1.91** | **36.13±1.81** | **94.26±2.02** |

Note: The deep learning methods are trained using all available training subjects. The means and standard deviations are computed over all test data, respectively. The lowest RLNE, highest PSNR and SSIM values are bold faced.

## VI. CONCLUSION

In this work, we present a new 1D learning scheme and further propose a One-dimensional Deep Low-rank and Sparse network (ODLS) for magnetic resonance image reconstruction in fast imaging. Specifically, the 1D learning scheme makes the deep network easier to train, especially suitable for the case of limited training subjects. We further design a compact 1D CNN architecture which unrolls the iteration procedure of a low-rank and sparse model into a deep network, to yield promising results. Extensive results on *in vivo* datasets demonstrate that, the proposed ODLS provides improved and more robust reconstruction performance than state-of-the-art methods both visually and quantitatively.

The success of the 1D learning scheme in fast magnetic resonance imaging is expected be useful for clinical applications, which only provide relatively small number of training subjects and limited computer memory.

## ACKNOWLEDGMENTS

The authors thank Xinlin Zhang and Jian Wu for assisting in data processing and helpful discussions. The authors thank Weiping He, Shaorong Fang, and Tianfu Wu from Information and Network Center of Xiamen University for the help with the GPU computing. The authors also thank Drs. Michael Lustig, Jong Chul Ye, Dosik Hwang, and Mathews Jacob for sharing their codes online.

## REFERENCES

[1] G. H. Mukesh, A. O'Shea, and R. Weissleder, "Advances in clinical MRI technology," *Sci. Transl. Med.*, vol. 11, no. 523, eaba2591, 2019.

[2] M. Lustig, D. Donoho, and J. M. Pauly, "Sparse MRI: The application of compressed sensing for rapid MR imaging," *Magn. Reson. Med.*, vol. 58, no. 6, pp. 1182-1195, 2007.

[3] K. P. Pruessmann, M. Weiger, M. B. Scheidegger, and P. Boesiger, "SENSE: Sensitivity encoding for fast MRI," *Magn. Reson. Med.*, vol. 42, no. 5, pp. 952-962, 1999.

[4] M. A. Griswold *et al.*, "Generalized autocalibrating partially parallel acquisitions (GRAPPA)," *Magn. Reson. Med.*, vol. 47, no. 6, pp. 1202-1210, 2002.

[5] Z. Liang, "Spatiotemporal imaging with partially separable functions," in *4th IEEE International Symposium on Biomedical Imaging (ISBI)*, 2007, pp. 988-991.

[6] M. Lustig and J. M. Pauly, "SPIRiT: Iterative self-consistent parallel imaging reconstruction from arbitrary k-space," *Magn. Reson. Med.*, vol. 64, no. 2, pp. 457-471, 2010.

[7] Y. Chen, X. Ye, and F. Huang, "A novel method and fast algorithm for MR image reconstruction with significantly under-sampled data," *Inverse Probl. Imaging*, vol. 4, no. 2, pp. 223-240, 2010.

[8] S. Ravishankar and Y. Bresler, "MR image reconstruction from highly undersampled k-space data by dictionary learning," *IEEE Trans. Med. Imaging*, vol. 30, no. 5, pp. 1028-1041, 2011.

[9] M. Guerquin-Kern, M. Haberlin, K. P. Pruessmann, and M. Unser, "A fast wavelet-based reconstruction method for magnetic resonance imaging," *IEEE Trans. Med. Imaging*, vol. 30, no. 9, pp. 1649-1660, 2011.

[10] X. Qu, Y. Hou, F. Lam, D. Guo, J. Zhong, and Z. Chen, "Magnetic resonance image reconstruction from undersampled measurements using a patch-based nonlocal operator," *Med. Image Anal.*, vol. 18, no. 6, pp. 843-856, 2014.

[11] Y. Yang, F. Liu, Z. Jin, and S. Crozier, "Aliasing artefact suppression in compressed sensing MRI for random phase-encode undersampling," *IEEE Trans. Biomed. Eng.*, vol. 62, no. 9, pp. 2215-2223, 2015.

[12] Y. Liu, Z. Zhan, J.-F. Cai, D. Guo, Z. Chen, and X. Qu, "Projected iterative soft-thresholding algorithm for tight frames in compressed sensing magnetic resonance imaging," *IEEE Trans. Med. Imaging*, vol. 35, no. 9, pp. 2130-2140, 2016.

[13] Z. Zhan, J.-F. Cai, D. Guo, Y. Liu, Z. Chen, and X. Qu, "Fast multiclass dictionaries learning with geometrical directions in MRI reconstruction," *IEEE Trans. Biomed. Eng.*, vol. 63, no. 9, pp. 1850-1861, 2016.

[14] X. Zhang *et al.*, "A guaranteed convergence analysis for the projected fast iterative soft-thresholding algorithm in parallel MRI," *Med. Image Anal.*, vol. 69, 101987, 2021.

[15] J. P. Haldar, "Autocalibrated loraks for fast constrained MRI reconstruction," in *IEEE 12th International Symposium on Biomedical Imaging (ISBI)*, 2015, pp. 910-913.

[16] X. Zhang *et al.*, "Image reconstruction with low-rankness and self-consistency of k-space data in parallel MRI," *Med. Image Anal.*, vol. 63, 101687, 2020.

[17] X. Zhang *et al.*, "Accelerated MRI reconstruction with separable and enhanced low-rank Hankel regularization," arXiv: 2107.11650, 2021.

[18] S. G. Lingala, Y. Hu, E. DiBella, and M. Jacob, "Accelerated dynamic MRI exploiting sparsity and low-rank structure: k-t SLR," *IEEE Trans. Med. Imaging*, vol. 30, no. 5, pp. 1042-1054, 2011.

[19] B. Zhao, J. P. Haldar, A. G. Christodoulou, and Z. Liang, "Image reconstruction from highly undersampled (k,t)-space data with joint partial separability and sparsity constraints," *IEEE Trans. Med. Imaging*, vol. 31, no. 9, pp. 1809-1820, 2012.

[20] R. Otazo, E. Candès, and D. K. Sodickson, "Low-rank plus sparse matrix decomposition for accelerated dynamic MRI with separation of background and dynamic components," *Magn. Reson. Med.*, vol. 73, no. 3, pp. 1125-1136, 2015.

[21] P. J. Shin *et al.*, "Calibrationless parallel imaging reconstruction based on structured low-rank matrix completion," *Magn. Reson. Med.*, vol. 72, no. 4, pp. 959-970, 2014.

[22] J. P. Haldar, "Low-rank modeling of local k-space neighborhoods (LORAKS) for constrained MRI," *IEEE Trans. Med. Imaging*, vol. 33, no. 3, pp. 668-681, 2014.

[23] K. H. Jin, D. Lee, and J. C. Ye, "A general framework for compressed sensing and parallel MRI using annihilating filter based low-rank Hankel matrix," *IEEE Trans. Comput. Imaging*, vol. 2, no. 4, pp. 480-495, 2016.

[24] G. Ongie and M. Jacob, "A fast algorithm for convolutional structured low-rank matrix recovery," *IEEE Trans. Comput. Imaging*, vol. 3, no. 4, pp. 535-550, 2017.

[25] Y. Hu, X. Liu, and M. Jacob, "A generalized structured low-rank matrix completion algorithm for MR image recovery," *IEEE Trans. Med. Imaging*, vol. 38, no. 8, pp. 1841-1851, 2019.

[26] S. Wang *et al.*, "Accelerating magnetic resonance imaging via deep learning," in *IEEE 13th International Symposium on Biomedical Imaging (ISBI)*, 2016, pp. 514-517.

[27] B. Zhu, J. Z. Liu, S. F. Cauley, B. R. Rosen, and M. S. Rosen, "Image reconstruction by domain-transform manifold learning," *Nature*, vol. 555, no. 7697, pp. 487-492, 2018.

[28] Y. Yang, J. Sun, H. Li, and Z. Xu, "ADMM-CSNet: A deep learning approach for image compressive sensing," *IEEE Trans. Pattern Anal. Mach. Intell.*, vol. 42, no. 3, pp. 521-538, 2020.

[29] D. Liang, J. Cheng, Z. Ke, and L. Ying, "Deep magnetic resonance image reconstruction: Inverse problems meet neural networks," *IEEE Signal Process. Mag.*, vol. 37, no. 1, pp. 141-151, 2020.

[30] J. C. Ye, Y. Han, and E. Cha, "Deep convolutional framelets: A general deep learning framework for inverse problems," *SIAM J. Imaging Sci.*, vol. 11, no. 2, pp. 991-1048, 2018.

[31] Y. Han, L. Sunwoo, and J. C. Ye, "k-space deep learning for accelerated MRI," *IEEE Trans. Med. Imaging*, vol. 39, no. 2, pp. 377-386, 2020.

[32] T. Eo, H. Shin, Y. Jun, T. Kim, and D. Hwang, "Accelerating Cartesian MRI by domain-transform manifold learning in phase-encoding direction," *Med. Image Anal.*, vol. 63, p. 101689, 2020.

[33] J. Zhang and B. Ghanem, "ISTA-Net: Interpretable optimization-inspired deep network for image compressive sensing," in *IEEE Conference on Computer Vision and Pattern Recognition (CVPR)*, 2018, pp. 1828-1837.

[34] T. Eo, Y. Jun, T. Kim, J. Jang, H.-J. Lee, and D. Hwang, "KIKI-net: Cross-domain convolutional neural networks for reconstructing undersampled magnetic resonance images," *Magn. Reson. Med.*, vol. 80, no. 5, pp. 2188-2201, 2018.

[35] K. Hammernik et al., "Learning a variational network for reconstruction of accelerated MRI data," *Magn. Reson. Med.*, vol. 79, no. 6, pp. 3055-3071, 2018.

[36] H. K. Aggarwal, M. P. Mani, and M. Jacob, "MoDL: Model-based deep learning architecture for inverse problems," *IEEE Trans. Med. Imaging*, vol. 38, no. 2, pp. 394-405, 2019.

[37] T. Lu *et al.*, "pFISTA-SENSE-ResNet for parallel MRI reconstruction," *J. Magn. Reson.*, vol. 318, 106790, 2020.

[38] M. Akçakaya, S. Moeller, S. Weingärtner, and K. Uğurbil, "Scan-specific robust artificial-neural-networks for k-space interpolation (RAKI) reconstruction: Database-free deep learning for fast imaging," *Magn. Reson. Med.*, vol. 81, no. 1, pp. 439-453, 2019.

[39] A. Pramanik, H. Aggarwal, and M. Jacob, "Deep generalization of structured low-rank algorithms (Deep-SLR)," *IEEE Trans. Med. Imaging*, vol. 39, no. 12, pp. 4186-4197, 2020.

[40] A. Sriram, J. Zbontar, T. Murrell, C. L. Zitnick, A. Defazio, and D. K. Sodickson, "GrappaNet: Combining parallel imaging with deep learning for multi-coil MRI reconstruction," in *IEEE Conference on Computer Vision and Pattern Recognition (CVPR)*, 2020, pp. 14303-14310.

[41] Z. Ke *et al.*, "Learned low-rank priors in dynamic MR imaging," *IEEE Trans. Med. Imaging*, pp. 1-1, 2021.

[42] D. C. Noll, D. G. Nishimura, and A. Macovski, "Homodyne detection in magnetic resonance imaging," *IEEE Trans. Med. Imaging*, vol. 10, no. 2, pp. 154-163, 1991.

[43] M. Jacob, M. P. Mani, and J. C. Ye, "Structured low-rank algorithms: Theory, magnetic resonance applications, and links to machine learning," *IEEE Signal Process. Mag.*, vol. 37, no. 1, pp. 54-68, 2020.

[44] J. Zhang, C. Liu, and M. E. Moseley, "Parallel reconstruction using null operations," *Magn. Reson. Med.*, vol. 66, no. 5, pp. 1241-1253, 2011.

[45] K. Mohan and M. Fazel, "Iterative reweighted algorithms for matrix rank minimization," vol. 13, no. 1, *J. Mach. Learn. Res.*, pp. 3441–3473, 2012.

[46] S. Ioffe and C. Szegedy, "Batch normalization: Accelerating deep network training by reducing internal covariate shift," arXiv: 1502.03167, 2015.

[47] X. Glorot and Y. Bengio, "Understanding the difficulty of training deep feedforward neural networks," in *International Conference on Artificial Intelligence and Statistics (AISTATS)*, 2010, pp. 249-256.

[48] M. Abadi *et al.*, "TensorFlow: Large-scale machine learning on heterogeneous distributed systems," arXiv: 1603.04467, 2016.

[49] T. Zhang, J. M. Pauly, S. S. Vasanawala, and M. Lustig, "Coil compression for accelerated imaging with Cartesian sampling," *Magn. Reson. Med.*, vol. 69, no. 2, pp. 571-582, 2013.

[50] W. Zhou, A. C. Bovik, H. R. Sheikh, and E. P. Simoncelli, "Image quality assessment: From error visibility to structural similarity," *IEEE Trans. Image Process.*, vol. 13, no. 4, pp. 600-612, 2004.

# "One-dimensional Deep Low-rank and Sparse Network for Accelerated MRI"

Zi Wang, Chen Qian, Di Guo, Hongwei Sun, Rushuai Li, Bo Zhao, and Xiaobo Qu*

## S1. Evaluation Criteria

The relative $l_2$ norm error (RLNE) is defined as

$$RLNE = \frac{\|\mathbf{x} - \hat{\mathbf{x}}\|_2}{\|\mathbf{x}\|_2}, \tag{S1-1}$$

where $\mathbf{x}$ and $\hat{\mathbf{x}}$ denote the column stacked fully sampled and reconstructed image, respectively. $\|\cdot\|_2$ is the $l_2$ norm. The lower RLNE indicates the lower reconstruction error.

The peak signal-to-noise ratio (PSNR) is defined as

$$PSNR(dB) = 10 \cdot \log_{10}\left(\frac{MN\|\mathbf{x}\|_\infty}{\|\mathbf{x} - \hat{\mathbf{x}}\|_2}\right), \tag{S1-2}$$

where $\mathbf{x}$ and $\hat{\mathbf{x}}$ denote the column stacked fully sampled and reconstructed image after combining by the square root of sum of squares, respectively. $\|\cdot\|_2$ is the $l_2$ norm, $\|\cdot\|_\infty$ is the infinite norm. $M$ and $N$ represent the dimension of the FE and PE, respectively. The higher PSNR indicates the less distortion in the reconstructed image.

The measure of structural similarity (SSIM) is defined as

$$SSIM = \frac{(2\mu_\mathbf{X}\mu_{\hat{\mathbf{X}}} + C_1)(2\sigma_{\mathbf{X}\hat{\mathbf{X}}} + C_2)}{(\mu_\mathbf{X}^2 + \mu_{\hat{\mathbf{X}}}^2 + C_1)(\sigma_\mathbf{X}^2 + \sigma_{\hat{\mathbf{X}}}^2 + C_2)}, \tag{S1-3}$$

where $\mathbf{X}$ and $\hat{\mathbf{X}}$ denote the fully sampled and reconstructed image after combining by the square root of sum of squares, respectively. $\mu_\mathbf{X}$, $\mu_{\hat{\mathbf{X}}}$, $\sigma_\mathbf{X}$, $\sigma_{\hat{\mathbf{X}}}$, $\sigma_{\mathbf{X}\hat{\mathbf{X}}}$ denote the means, standard deviations, and covariance of $\mathbf{X}$ and $\hat{\mathbf{X}}$, respectively. Constant $C_1$ and $C_2$ are introduced to avoid the case when the denominator multiplied by $\mu_\mathbf{X}^2 + \mu_{\hat{\mathbf{X}}}^2$ and $\sigma_\mathbf{X}^2 + \sigma_{\hat{\mathbf{X}}}^2$ is close to zero. The higher SSIM indicates the better detail preservation in the reconstruction.

# S2. Detailed Results for Matched Reconstruction

**Table S2-1. RLNE (×10⁻²)/PSNR (dB)/SSIM (×10⁻²) of reconstructions using different number of training subjects $N_{TS}$ of a knee dataset (mean±std).**

| Method | $N_{TS}$ | RLNE | PSNR | SSIM |
|---|---|---|---|---|
| GRAPPA | / | 15.46±2.13 | 30.44±1.31 | 83.27±2.59 |
| IUNET | 2 | 23.07±4.17 | 27.23±2.32 | 79.42±3.63 |
|  | 4 | 16.87±2.13 | 29.35±1.41 | 82.98±2.81 |
|  | 8 | 14.69±2.04 | 30.27±1.49 | 84.49±2.56 |
|  | 12 | 14.53±1.68 | 30.62±1.40 | 84.91±2.28 |
| DOTA | 2 | 18.60±3.35 | 28.28±1.89 | 80.39±4.51 |
|  | 4 | 15.15±2.45 | 30.26±1.65 | 84.38±3.23 |
|  | 8 | 13.57±2.29 | 31.40±1.74 | 85.63±2.77 |
|  | 12 | 12.32±2.05 | 32.35±1.64 | 87.29±2.17 |
| HDSLR | 2 | 17.24±3.40 | 28.62±2.02 | 81.03±4.18 |
|  | 4 | 13.76±2.48 | 30.76±1.84 | 84.99±3.12 |
|  | 8 | 12.06±2.20 | 31.99±1.81 | 86.65±2.56 |
|  | 12 | 10.91±1.77 | 33.03±1.78 | 88.43±2.15 |
| ODLS | 2 | 14.71±2.68 | 30.01±1.96 | 83.93±3.57 |
|  | 4 | 11.87±1.91 | 32.07±1.79 | 87.36±2.47 |
|  | 8 | 10.46±1.71 | 33.37±1.77 | 89.06±2.14 |
|  | 12 | 9.78±1.54 | 34.15±1.81 | 90.02±2.01 |

Note: The Cartesian undersampling pattern with AF=4 is used. "/" represents that the corresponding method requires no training. The means and standard deviations are computed over all test data, respectively.

**Table S2-2. RLNE (×10$^{-2}$)/PSNR (dB)/SSIM (×10$^{-2}$) of reconstructions using different number of training subjects $N_{TS}$ of a brain dataset (mean±std).**

| Method | $N_{TS}$ | RLNE | PSNR | SSIM |
|--------|----------|------|------|------|
| GRAPPA | / | 14.92±1.75 | 32.95±1.66 | 91.16±2.60 |
| IUNET | 1 | 23.22±3.09 | 29.12±1.55 | 88.14±2.69 |
| | 2 | 17.64±2.26 | 31.51±1.62 | 89.15±2.55 |
| | 4 | 16.00±2.19 | 31.83±1.56 | 90.81±2.79 |
| | 6 | 15.34±2.09 | 32.87±1.56 | 91.46±2.59 |
| DOTA | 1 | 16.96±2.12 | 31.85±1.52 | 90.67±2.26 |
| | 2 | 15.33±1.93 | 32.73±1.52 | 91.80±2.23 |
| | 4 | 13.38±1.93 | 33.93±1.61 | 92.97±2.12 |
| | 6 | 12.78±1.97 | 34.34±1.69 | 93.47±2.04 |
| HDSLR | 1 | 17.10±3.01 | 31.85±1.81 | 90.55±2.39 |
| | 2 | 14.76±2.51 | 33.11±1.77 | 91.60±2.43 |
| | 4 | 12.77±2.09 | 34.36±1.73 | 92.85±2.20 |
| | 6 | 11.93±1.95 | 34.95±1.84 | 93.29±2.19 |
| ODLS | 1 | 13.82±1.87 | 33.63±1.61 | 91.84±2.47 |
| | 2 | 12.62±1.84 | 34.44±1.63 | 92.67±2.31 |
| | 4 | 11.54±1.87 | 35.24±1.76 | 93.47±2.21 |
| | 6 | 10.94±1.82 | 35.71±1.85 | 93.85±2.17 |

Note: The Cartesian undersampling pattern with AF=4 is used. The means and standard deviations are computed over all test data.

# S3. More Reconstruction Results under Different Undersampling Scenarios

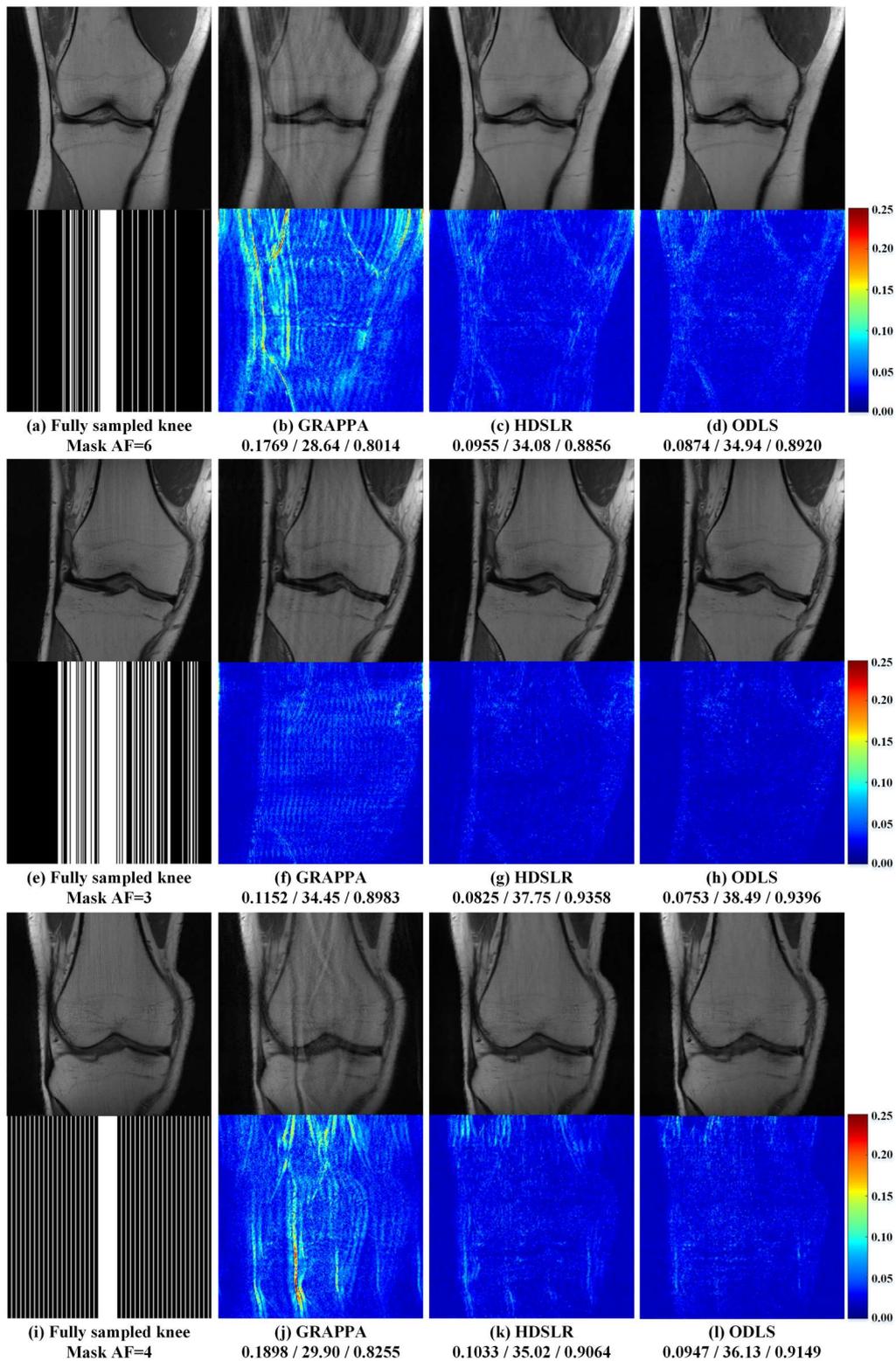

**Fig. S3-1.** Reconstruction results of a knee dataset under different undersampling scenarios. (a)-(d) are reconstructions using the Cartesian undersampling pattern with AF=6. (e)-(h) are reconstructions using the partial Fourier undersampling pattern with AF=3. (i)-(l) are reconstructions using the uniform undersampling pattern with AF=4. Note: RLNE/PSNR(dB)/SSIM are listed for each reconstruction.

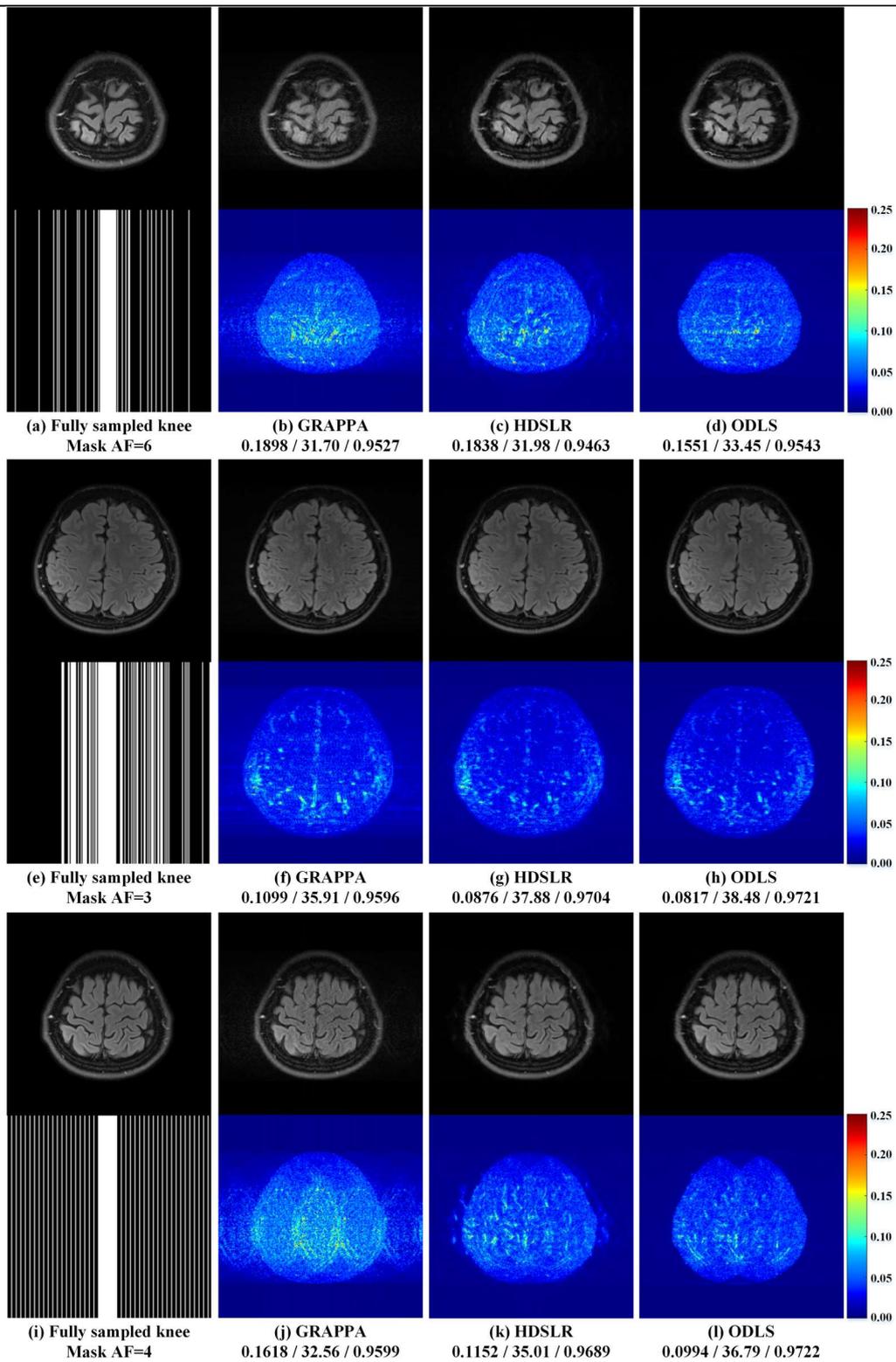

**Fig. S3-2.** Reconstruction results of a brain dataset under different undersampling scenarios. (a)-(d) are reconstructions using the Cartesian undersampling pattern with AF=6. (e)-(h) are reconstructions using the partial Fourier undersampling pattern with AF=3. (i)-(l) are reconstructions using the uniform undersampling pattern with AF=4. Note: RLNE/PSNR(dB)/SSIM are listed for each reconstruction.

# S4. Parameters of GRAPPA

We ran reconstruction experiments using GRAPPA with a series of combinations of model-specific parameters subjecting to each dataset, and the parameters of GRAPPA allowing the lowest RLNEs were chosen. The detailed parameter settings of GRAPPA are listed in TABLE S5-1.

**Table S4-1. Parameters of GRAPPA.**

| Dataset | Pattern | Kernel size | Kernel calibration parameter |
|---|---|---|---|
| Coronal proton density weighted knee | Cartesian AF=4 | 7×7 | 0.05 |
| | Cartesian AF=6 | 7×7 | 0.10 |
| | 3/4 partial Fourier AF=3 | 7×7 | 0.10 |
| | Uniform AF=4 | 7×7 | 0.10 |
| Sagittal proton density weighted knee | Cartesian AF=4 | 7×7 | 0.05 |
| | Cartesian AF=6 | 7×7 | 0.10 |
| | 3/4 partial Fourier AF=3 | 7×7 | 0.10 |
| | Uniform AF=4 | 7×7 | 0.10 |
| Axial $T_2$ weighted brain | Cartesian AF=4 | 5×5 | 0.05 |
| | Cartesian AF=6 | 5×5 | 0.10 |
| | 3/4 partial Fourier AF=3 | 5×5 | 0.10 |
| | Uniform AF=4 | 5×5 | 0.10 |
| Axial $T_1$ weighted brain | Cartesian AF=4 | 5×5 | 0.05 |
| | Cartesian AF=6 | 5×5 | 0.10 |
| | 3/4 partial Fourier AF=3 | 5×5 | 0.10 |
| | Uniform AF=4 | 5×5 | 0.10 |